\documentclass[aps,prl,twocolumn,preprintnumbers,superscriptaddress]{revtex4-2}
\usepackage{graphicx}
\usepackage{physics}
\usepackage{amsmath}
\usepackage{amssymb}
\usepackage{bm}
\usepackage{color}
\definecolor{LinkColor}{rgb}{0.0,0.0,1}
\usepackage{hyperref}
\hypersetup{
	colorlinks=true,
	citecolor=LinkColor,
	linkcolor=LinkColor,
	urlcolor=LinkColor,
}

\begin{document}

\title{Chiral spin liquid and chiral antiferromagnetism in half-filled moiré Hubbard model: possible applications to twisted bilayer TMDs}

\author{Chuyi Tuo}
\affiliation{Institute for Advanced Study, Tsinghua University, Beijing 100084, China}

\author{Hong Yao}
\email{yaohong@tsinghua.edu.cn}
\affiliation{Institute for Advanced Study, Tsinghua University, Beijing 100084, China}

\date{\today}

\begin{abstract}
Twisted transition metal dichalcogenides offer an exceptionally tunable moiré platform for studying correlation physics beyond conventional condensed matter systems. In particular, the intriguing interplay between the displacement field and the twist angle remains to be fully resolved. In this paper, we use large-scale density matrix renormalization group simulations to study the minimal moiré Hubbard model on a triangular lattice at half-filling, where the displacement field effect is captured by a spin-dependent staggered flux. We find that the displacement field significantly enriches the triangular Hubbard phase diagram in several qualitative ways. It rapidly destabilizes the chiral spin liquid phase beyond a narrow weak-field regime, induces pronounced chiral correlations in the strong-coupling $120^\circ$-antiferromagnetic phase, and stabilizes incommensurate spin-density wave phases at weaker coupling. We further find signatures of a continuous transition between the chiral spin liquid and chiral antiferromagnetic phases at a finite displacement field, potentially driven by spinon condensation. Our results uncover rich displacement-field-driven many-body physics and provide useful guidance for future experiments in moiré superlattice systems.
\end{abstract}

\maketitle

\textit{Introduction.}---
Moiré superlattice systems~\cite{bistritzerMoireBandsTwisted2011,andreiGrapheneBilayersTwist2020,balentsSuperconductivityStrongCorrelations2020a,andreiMarvelsMoireMaterials2021,kennesMoireHeterostructuresCondensedmatter2021a,castellanos-gomezVanWaalsHeterostructures2022,makSemiconductorMoireMaterials2022,nuckollsMicroscopicPerspectiveMoire2024} have emerged as versatile experimental platforms for realizing strongly correlated electronic states, with twist angle, displacement field, and carrier density offering precise control over their electronic properties. In particular, twisted transition metal dichalcogenides (TMDs) constitute a key class of moiré systems that can host a wide variety of exotic phases~\cite{liQuantumPhasesTwisted2026}, including correlated insulators~\cite{tangSimulationHubbardModel2020,reganMottGeneralizedWigner2020,wangCorrelatedElectronicPhases2020,xuCorrelatedInsulatingStates2020,liContinuousMottTransition2021,ghiottoQuantumCriticalityTwisted2021,xuTunableBilayerHubbard2022}, integer and fractional quantum anomalous Hall~\cite{liQuantumAnomalousHall2021,caiSignaturesFractionalQuantum2023,zengThermodynamicEvidenceFractional2023,parkObservationFractionallyQuantized2023,xuObservationIntegerFractional2023a} and quantum spin Hall~\cite{kangEvidenceFractionalQuantum2024,xuInterplayTopologyCorrelations2025} states, composite Fermi liquids~\cite{parkObservationFractionallyQuantized2023,andersonTrionSensingZerofield2024a}, as well as superconductivity~\cite{xiaSuperconductivityTwistedBilayer2025,guoSuperconductivity50degTwisted2025,xiaBandwidthtunedMottTransition2026,xuSignaturesUnconventionalSuperconductivity2026}. For a broad class of moiré TMD systems, the low-energy physics can be effectively captured by a Hubbard model~\cite{wuHubbardModelPhysics2018,wuTopologicalInsulatorsTwisted2019} with distinctive moiré features. Understanding Hubbard physics in this moiré setting is therefore crucial for elucidating the microscopic origin of the correlated phases observed in moiré TMDs, and for identifying routes toward realizing novel quantum phases.

\begin{figure}[b]
    \centering
    \includegraphics[width=\linewidth]{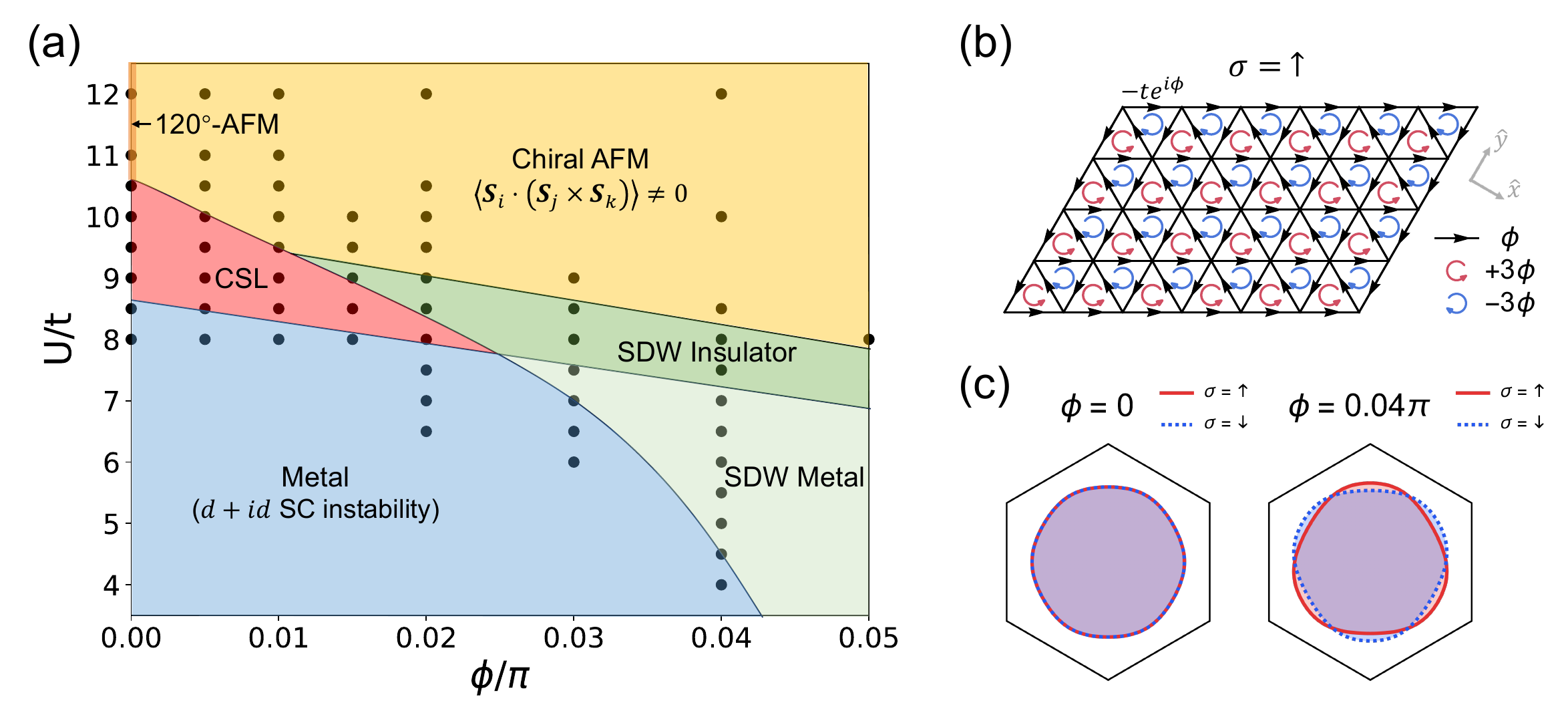}
    \caption{(a) Ground state phase diagram of the half-filled moiré Hubbard model in the strongly correlated regime, obtained by DMRG simulations on YC-$48\times4$ cylinders.  Black dots denote the simulated data points. The CSL to chiral AFM transition appears to be continuous within our numerical resolution. Additional $L_y=6$ simulations are performed to assess finite-circumference effects. (b) Schematic illustration of the staggered hopping flux pattern for spin up ($\sigma=\uparrow$); the flux pattern is reversed for spin down ($\sigma=\downarrow$) by time-reversal symmetry $\mathcal{T}$. (c) Half-filled Fermi surfaces for both spins $\sigma=\uparrow,\downarrow$ at two representative displacement fields $\phi=0$ and $\phi=0.04\pi$, respectively. }
    \label{Fig1}
\end{figure}

Among these moiré specific features, the displacement field plays a particularly important role, coupling directly to layer degrees of freedom through an interlayer energy imbalance. 
Although the displacement field is spin independent at the microscopic level, upon projecting onto a spin-valley locked~\cite{xiaoCoupledSpinValley2012} miniband of twisted bilayer TMDs, it can manifest as an effective spin-orbit coupling (SOC) that strongly modifies the low-energy electronic structure. This motivates a minimal moiré Hubbard model description~\cite{panBandTopologyHubbard2020,zangHartreeFockStudyMoire2021,zangDynamicalMeanFieldTheory2022b,wietekTunableStripeOrder2022,wuPairDensityWaveChiralSuperconductivity2023,biborskiSignaturesSuperconductingPairing2025} on triangular lattice for twisted TMD homobilayers, where the displacement-field-induced SOC is captured by a tunable spin-dependent staggered flux, and the relative strength of interaction and hopping is controlled by the twist angle. This model serves as a starting point for understanding the many-body physics in twisted TMD homobilayers, and provides a fundamental framework for clarifying the interplay between SOC and strong correlations.
Despite substantial theoretical and numerical progress on moiré Hubbard~\cite{panBandTopologyHubbard2020,zangHartreeFockStudyMoire2021,zangDynamicalMeanFieldTheory2022b,wietekTunableStripeOrder2022,wuPairDensityWaveChiralSuperconductivity2023,biborskiSignaturesSuperconductingPairing2025} and related models~\cite{panQuantumPhaseDiagram2020,devakulMagicTwistedTransition2021,liSpontaneousFractionalChern2021,zhouQuantumPhasesTransition2022,morales-duranNonlocalInteractionsMoire2022,belangerSuperconductivityTwistedBilayer2022,kieseTMDsPlatformSpin2022,chenSingletTripletPair2023,zhouChiralNodalSuperconductors2023,zegrodnikMixedSinglettripletSuperconducting2023,kleblCompetitionDensityWaves2023,crepelTopologicalSuperconductivityDoped2023,qiuInteractiondrivenTopologicalPhase2023,reddyFractionalQuantumAnomalous2023,dongCompositeFermiLiquid2023,goldmanZerofieldCompositeFermi2023,reddyGlobalPhaseDiagram2023a,xuMaximallyLocalizedWannier2024,crepelBridgingSmallLarge2024,akbarTopologicalSuperconductivityMixed2024,schradeNematicChiralTopological2024a,wangFractionalChernInsulator2024,kimTheoryCorrelatedInsulators2025,myerson-jainSuperconductorInsulatorTransitionTMD2024,zhuSuperconductivityTwistedTransition2025,christosApproximateSymmetriesInsulators2025a,xieSuperconductivityTwistedWSe2025,guerciTopologicalSuperconductivityRepulsive2024a,tuoTheoryTopologicalSuperconductivity2025,qinTopologicalChiralSuperconductivity2025,fischerTheoryIntervalleycoherentAFM2025,jinGossamerSuperconductivityMoire2026,sharmaTopologicalQuantumPhase2024,jiaMoireFractionalChern2024,yuFractionalChernInsulators2024,reddyNonAbelianFractionalizationTopological2024,ahnNonAbelianFractionalQuantum2024,wangHigherLandauLevelAnalogs2025,xuMultipleChernBands2025,chenRobustNonAbelianEvendenominator2025,shenMagnetorotonsMoireFractional2026,chenFractionalChernInsulator2026,heFractionalChernInsulators2025,tuoFractionalQuantumAnomalous2025,divicAnyonSuperconductivityTopological2025,kuhlenkampRobustSuperconductivityDoping2025,chenTopologicalChiralSuperconductivity2026,guerciTopologicalSuperconductivityEmergent2026,fanHiddenWeakpairingSuperconductivity2026}, a comprehensive many-body understanding of displacement field effects in the strongly correlated regime remains limited, especially compared with the well-studied zero-field triangular lattice cases~\cite{arovasHubbardModel2022,qinHubbardModelComputational2022,szaszChiralSpinLiquid2020,chenQuantumSpinLiquid2022,zhuChiralSpinLiquid2024,szaszPhaseDiagramAnisotropic2021,sunSinglebandTriangularLattice2026,zhuDopedMottInsulators2022,capriottiLongRangeNeelOrder1999,whiteNeelOrderSquare2007a,shirakawaGroundstatePhaseDiagram2017,wietekMottInsulatingStates2021,chenGroundStatesSpin2013,laubachPhaseDiagramHubbard2015,huCompetingSpinliquidStates2015,gongGlobalPhaseDiagram2017,gongChiralSpinLiquid2019,huDiracSpinLiquid2019,venderleyDensityMatrixRenormalization2019a,jiangTopologicalSuperconductivityDoped2020,pengGaplessSpinLiquid2021,jiangSuperconductivityDopedQuantum2021,jiangNatureQuantumSpin2023}.
In particular, increasing numerical evidence~\cite{szaszChiralSpinLiquid2020,chenQuantumSpinLiquid2022,zhuChiralSpinLiquid2024,szaszPhaseDiagramAnisotropic2021,sunSinglebandTriangularLattice2026} suggests that, in the absence of displacement field, the pure triangular lattice Hubbard model hosts a chiral spin liquid (CSL)~\cite{kalmeyerEquivalenceResonatingvalencebondFractional1987,wenChiralSpinStates1989,schroeterSpinHamiltonianWhich2007,yaoExactChiralSpin2007} phase between the metallic and $120^\circ$-antiferromagnetic (AFM) Mott insulating phases. 
How these phases evolve and whether additional phases may emerge under a finite displacement field remain important open questions.

In this paper, we employ large-scale density matrix renormalization group (DMRG)~\cite{whiteDensityMatrixFormulation1992,whiteDensitymatrixAlgorithmsQuantum1993,schollwockDensitymatrixRenormalizationGroup2005,schollwockDensitymatrixRenormalizationGroup2011} simulations to map out the displacement-field-dependent ground state phase diagram of the moiré Hubbard model at half-filling in the strongly correlated regime. The resulting DMRG phase diagram on $L_y=4$ cylinders is illustrated in Fig.~\ref{Fig1}(a). 
We find that the displacement field substantially reshapes the triangular Hubbard phase diagram in the following three key aspects. 
First, our results suggest that the CSL phase is rapidly suppressed with increasing displacement field and survives only up to $\phi \sim 0.02\pi$. 
Second, while the displacement field is known to favor $120^\circ$-AFM order in the $xy$-plane at strong coupling~\cite{panBandTopologyHubbard2020,zangHartreeFockStudyMoire2021,zangDynamicalMeanFieldTheory2022b,wietekTunableStripeOrder2022,wuPairDensityWaveChiralSuperconductivity2023,biborskiSignaturesSuperconductingPairing2025}, we demonstrate that it also stabilizes long-range scalar chirality order, giving rise to a chiral AFM phase.
The resulting CSL to chiral AFM transition appears to be continuous, potentially driven by spinon condensation, offering a promising route towards fractionalized quantum criticality~\cite{sachdevKagomeTriangularlatticeHeisenberg1992,wangSpinliquidStatesTriangular2006,senthilDeconfinedQuantumCritical} with chiral order persisting on both sides of the transition.
Third, at finite displacement field, we identify incommensurate spin-density wave (IC-SDW) metal and insulator phases in the intermediate-coupling regime.
Taken together, these results establish a concrete theoretical foundation on future experimental searches for unconventional phases and phase transitions in moiré TMD systems.

\textit{Model and method.}---
We consider the minimal moiré Hubbard model on triangular lattice with nearest-neighbor hopping and onsite Hubbard repulsion~\cite{panBandTopologyHubbard2020,zangHartreeFockStudyMoire2021,zangDynamicalMeanFieldTheory2022b,wietekTunableStripeOrder2022,wuPairDensityWaveChiralSuperconductivity2023,biborskiSignaturesSuperconductingPairing2025}:
\begin{equation}
    H = -\sum_{\langle i,j\rangle, \sigma} \left( t e^{i\sigma\phi_{ij}} c^\dagger_{i\sigma} c_{j\sigma} + \text{h.c}.\right) + U \sum_i n_{i\uparrow} n_{i\downarrow}
\end{equation}
where $t$ is the hopping amplitude, $U$ is the onsite Hubbard repulsion, $\sigma=\uparrow,\downarrow$ (or $\pm1$) labels the spin-valley locked index~\cite{xiaoCoupledSpinValley2012}, and $c^\dagger_{i\sigma}, n_{i\sigma}$ denote the fermion creation and density operator at site $i$ with spin-valley $\sigma$. In the following, we refer to the spin-valley-locked index $\sigma$ simply as spin for brevity.

The most distinctive feature of the moiré Hubbard model is the nontrivial hopping phase factor $e^{i\sigma\phi_{ij}}$, generated by the displacement field that breaks symmetries between two layers, reducing the point group symmetry from $D_{6h}$ to $C_{3v}$ and the spin symmetry from $\mathrm{SU}(2)$ to $\mathrm{U}(1)$. The magnitude of the phase $\phi$ increases monotonically with the applied displacement field, and experimentally accessible values are estimated to satisfy $|\phi|\lesssim \pi/3$~\cite{zangHartreeFockStudyMoire2021}.  The hopping phases $\phi_{ij}$ and the resulting flux configuration for the $\sigma=\uparrow$ sector are illustrated in Fig.~\ref{Fig1}(b), where the hopping along the arrow direction corresponds to $\phi_{ij}=+\phi$, while the opposite direction yields $\phi_{ij}=-\phi$. Consequently, the two inequivalent elementary triangular plaquettes accumulate opposite fluxes, $\pm 3\phi$ respectively, resulting in zero net flux per moiré unit cell. Time-reversal symmetry $\mathcal{T}$ enforces that the $\sigma=\downarrow$ sector experiences the opposite flux pattern. As a result, the non-interacting Fermi surfaces for $\sigma=\uparrow,\downarrow$ are split but remain related by $\mathcal{T}$ at finite displacement field, as shown in Fig.~\ref{Fig1}(c).

To explore the many-body physics of the moiré Hubbard model at half-filling, with a focus on the displacement field effects in strongly correlated regime, we employ large-scale DMRG simulations, which are particularly effective for frustrated fermionic systems on finite width cylinders. 
We adopt YC cylinder geometry for triangular lattice, focusing primarily on systems with circumference $L_y=4$ and a large length $L_x$ up to $48$ to reduce open boundary effects. 
Our DMRG simulations explicitly preserve $U(1)_\text{charge}\times U(1)_\text{spin}$ symmetry, with bond dimensions up to $D=10000$, achieving typical truncation errors of order $10^{-5}$ to $10^{-7}$. We also perform preliminary simulations on $L_y=6$ systems to examine finite circumference effects~\cite{supp}. These simulations enable us to reliably characterize the ground state properties of the model in the strongly correlated regime of interest.

\begin{figure}[t]
    \centering
    \includegraphics[width=0.6\linewidth]{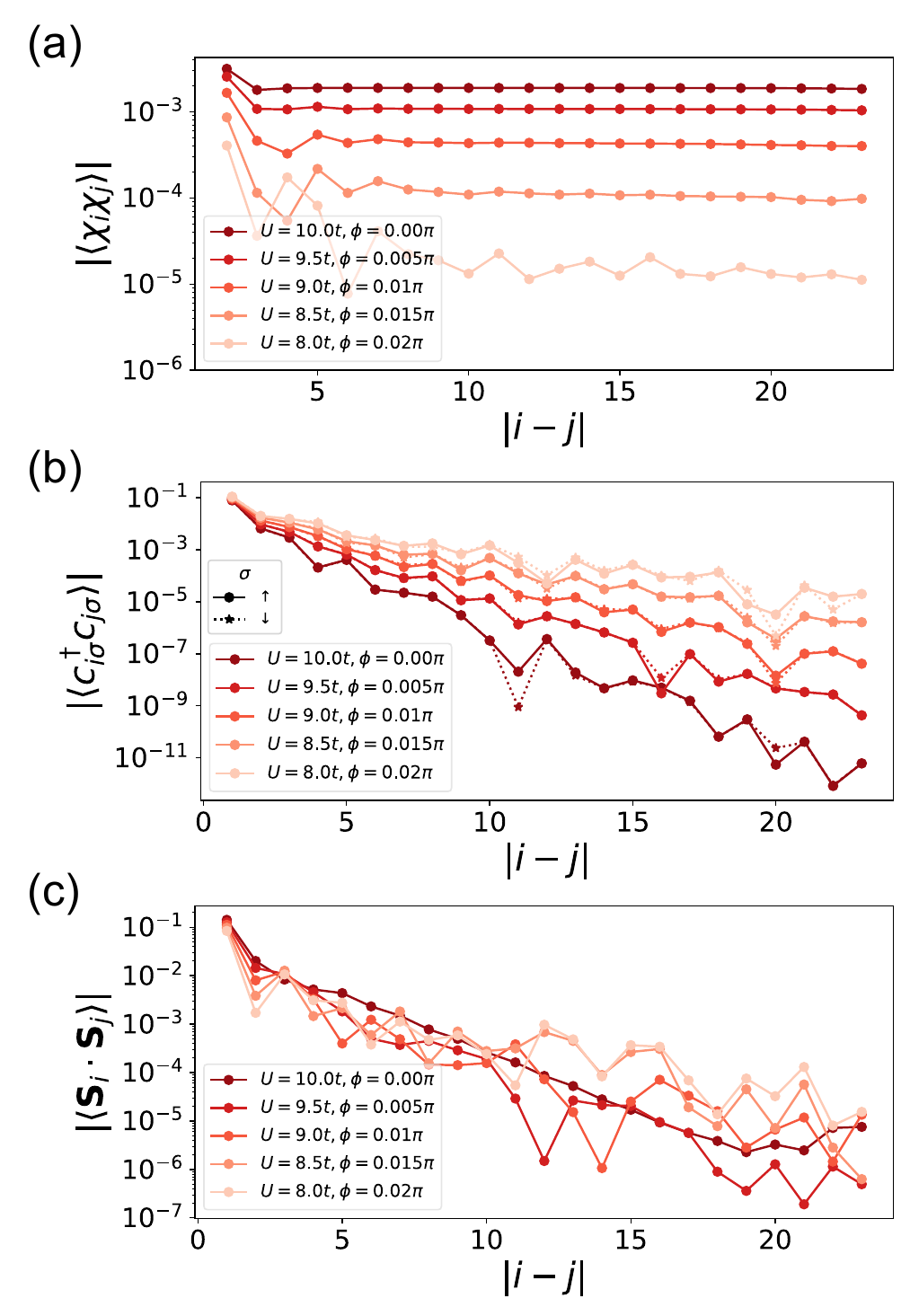}
    \caption{Numerical evidence of the CSL phase on $L_y=4$ cylinders along the parameter line cut indicated in the legends. (a) Scalar chirality correlation $\langle \chi_i \chi_j \rangle$. (b) Single-particle Green's function $\langle c^\dagger_{i\sigma} c_{j\sigma} \rangle$. (c) Spin correlation function $\langle \mathbf{S}_i \cdot \mathbf{S}_j \rangle$.}
    \label{Fig2}
\end{figure}

\textit{Chiral spin liquid.}---
The CSL phase in the zero-field triangular lattice Hubbard model has been substantially established~\cite{szaszChiralSpinLiquid2020,chenQuantumSpinLiquid2022,zhuChiralSpinLiquid2024,szaszPhaseDiagramAnisotropic2021,sunSinglebandTriangularLattice2026}, characterized by non-magnetic insulating behavior with long-range scalar chirality order, fractionally quantized spin Hall response, and characteristic entanglement spectrum counting. 
At zero displacement field, our results on $L_y=4$ cylinders are generally consistent with previous studies, which identified a CSL phase between the metallic and $120^\circ$-AFM phases for $9 \lesssim U/t \lesssim 10.5$~\cite{szaszChiralSpinLiquid2020,chenQuantumSpinLiquid2022,zhuChiralSpinLiquid2024}.
We now focus on the fate of this CSL phase under a finite displacement field $\phi$, following a representative line cut where $U/t$ is gradually reduced as $\phi$ increases, chosen to track the region where the CSL is relatively strong.
We first characterize the CSL by measuring the scalar chirality correlation function $\langle \chi_i \chi_j \rangle$, with the scalar chirality defined as:
\begin{equation}
\chi_i = \bm{S}_i \cdot \left(\bm{S}_{i+\bm{e}_1} \times \bm{S}_{i+\bm{e}_2}\right)
\end{equation}
where $(i, i+\bm{e}_1, i+\bm{e}_2)$ are sites on the same elementary triangle plaquette, and $\bm{S}_i$ is the spin operator at site $i$.
As shown in Fig.~\ref{Fig2}(a), the scalar chirality correlation functions remain finite at long distances along the chosen parameter line cut, indicating long-range chiral order consistent with the CSL phase. Fig.~\ref{Fig2}(b) shows that the single particle Green's functions $\langle c_{i\sigma}^\dagger c_{j\sigma}\rangle$ decay exponentially, suggesting a single particle gapped insulating behavior. The rapid exponential decay of the spin correlation functions $\langle \bm S_i \cdot \bm S_j\rangle$ in Fig.~\ref{Fig2}(c) further rules out magnetic long-range order in this regime. 
Our correlation measurements, together with adiabatic continuity from the more firmly established zero-field CSL and the expected stability of a fully gapped topological phase, strongly support interpreting this regime as finite field continuation of the zero-field CSL phase.

However, the CSL phase is clearly weakened as the displacement field $\phi$ increases, as evidenced by the suppression of chiral order (Fig.~\ref{Fig2}(a)), the weakening of insulating behavior (Fig.~\ref{Fig2}(b)), and the shrinking CSL reagion at larger $\phi$ (Fig~\ref{Fig1}(a)).
This trend is consistent with the physical expectation that displacement-field-induced SOC lowers the spin symmetry from $\mathrm{SU}(2)$ to $\mathrm{U}(1)$, suppressing quantum fluctuations that help to stabilize the CSL.
Our results provide a quantitative estimate of the stability of the CSL phase against finite displacement field, showing that it survives only up to $\phi\sim 0.02\pi$, which corresponds to interlayer energy difference $V_z\sim 3$ meV for $4^\circ$ twisted bilayer WSe$_2$ according to Ref.~\cite{panBandTopologyHubbard2020}, or $D_z$ of order $10 \,\text{mV}\cdot\text{nm}^{-1}$ estimated using typical device parameters of Ref.~\cite{xiaSuperconductivityTwistedBilayer2025}. This suggests that maintaining a sufficiently small displacement field is crucial for stabilizing the CSL phase in twisted TMD homobilayers.

\textit{Strong-coupling chiral antiferromagnetism.}---
We now turn to the larger $U/t$ regime and examine how the displacement field modifies the magnetic structure of the usual $120^\circ$-AFM phase.
From the theoretical perspective, a natural starting point is the strong-coupling expansion of the moiré Hubbard model~\cite{panBandTopologyHubbard2020,zangHartreeFockStudyMoire2021}, which takes the form of a displacement-field-tunable easy-axis XXZ anisotropy and Dzyaloshinskii-Moriya (DM) interaction:
\begin{equation}\label{eq:strongcoupling}
\begin{split}
H = \sum_{\langle ij\rangle} \Bigl[
& J S_i^z S_j^z
+ J\cos (2\phi) \left(S_i^x S_j^x + S_i^y S_j^y\right) \\
& + J\sin (2\phi) \, \hat{\bm z}\cdot
\left(\bm S_i \times \bm S_j\right)
\Bigr]
\end{split}
\end{equation}
where the superexchange coupling $J=4t^2/U$, and the bond orientation $(i,j)$ in the DM term follows the arrow direction of Fig.~\ref{Fig1}(b).
At $\phi=0$, Eq.~\eqref{eq:strongcoupling} reduces to the usual isotropic Heisenberg model, whose ground state is the $120^\circ$-AFM with an arbitrary ordering plane and two degenerate vector chiralities $\bm{\kappa}\sim\sum_{(i,j)\in\triangle }\bm{S}_i\times\bm{S}_j$, associated with $\mathrm{SU}(2)$ and inversion symmetries, respectively. Both of these symmetries are explicitly broken by a finite displacement field $\phi$. 
For sufficiently small $\phi$, the DM term dominates the magnetic anisotropy, naturally favoring $xy$-plane $120^\circ$-AFM order with fixed vector chirality~\cite{panBandTopologyHubbard2020,zangHartreeFockStudyMoire2021,zangDynamicalMeanFieldTheory2022b,wietekTunableStripeOrder2022,wuPairDensityWaveChiralSuperconductivity2023,biborskiSignaturesSuperconductingPairing2025}.
The subleading easy-axis XXZ anisotropy nevertheless provides a weak competing tendency toward a Y-state spin supersolid~\cite{leggettCanSolidBe1970,wesselSupersolidHardCoreBosons2005,heidarianPersistentSupersolidPhase2005,melkoSupersolidOrderDisorder2005,boninsegniSupersolidPhaseHardCore2005,wangExtendedSupersolidPhase2009,jiangSupersolidOrderFrustrated2009,heidarianSupersolidityTriangularLattice2010,yamamotoQuantumPhaseDiagram2014,sellmannPhaseDiagramAntiferromagnetic2015a,gallegosPhaseDiagramEasyAxis2025,kadosawaNontrivialThreesublatticeMagnetization2026}.
Moving beyond the strong-coupling limit Eq.~\eqref{eq:strongcoupling}, the situation is further complicated by higher-order virtual hopping processes and enhanced quantum fluctuations in the moiré Hubbard model, potentially enriching the ground state behavior. These effects motivate a direct numerical examination of the AFM phase, particularly in the strongly correlated regime.

\begin{figure}[t]
    \centering
    \includegraphics[width=\linewidth]{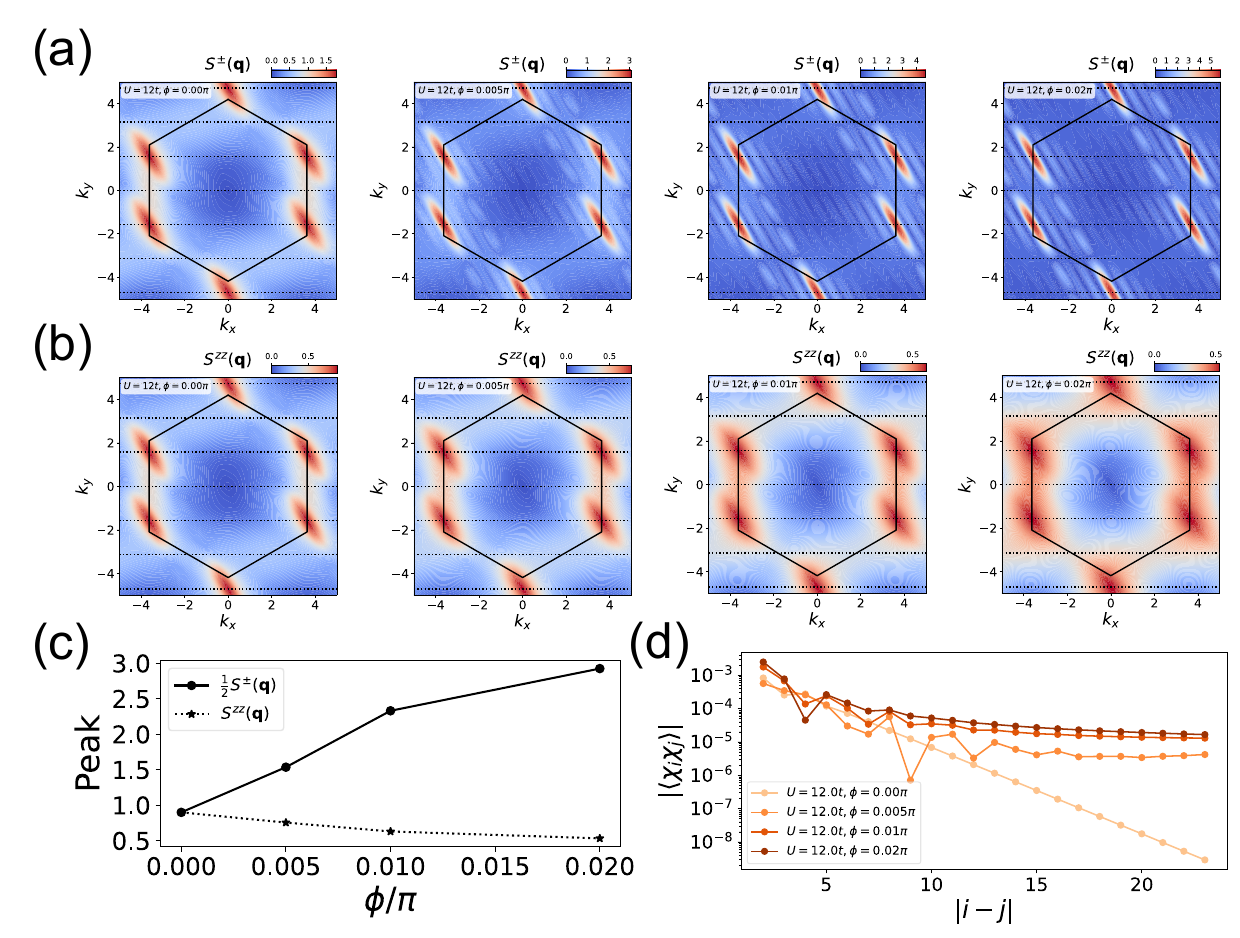}
    \caption{Characterization of the strong-coupling chiral AFM phase at $U/t=12$ for different displacement fields $\phi$ on $L_y=4$ cylinders. (a, b) Spin structure factors $S^{+-}(\mathbf{q})$ and $S^{zz}(\mathbf{q})$. (c) Peak values of $S^{+-}(\mathbf{q})$ and $S^{zz}(\mathbf{q})$ as a function of $\phi$. (d) Scalar chirality correlation $\langle \chi_i \chi_j \rangle$.}
    \label{Fig3}
\end{figure}

Fig.~\ref{Fig3} presents our DMRG characterization of the magnetic properties at $U/t=12$ for relatively small $\phi$ on $L_y=4$ cylinders. 
Fig.~\ref{Fig3}(a) and (b) show the static spin structure factors $S^{+-}(\bm q)$ and $S^{zz}(\bm q)$, defined as the spatial Fourier transform of the spin correlation functions:  
\begin{equation}
    S^{\alpha\beta}(\bm q) = \sum_j e^{i\bm q\cdot(\bm r_j-\bm r_i)}\langle S_i^\alpha S_j^\beta\rangle
\end{equation}
where $\alpha\beta=+-,zz$ denote the in-plane and out-of-plane channels, $i$ is a reference site chosen near the center of the system, and the summation over $j$ is restricted to the central half of the system to reduce open boundary effects. 
Although the YC4 cylinder does not perfectly accommodate the $K$ points, the dominant peaks here occur at the closest available momenta, consistent with previous studies~\cite{szaszChiralSpinLiquid2020,chenQuantumSpinLiquid2022,zhuChiralSpinLiquid2024}. With increasing $\phi$, the $S^{+-}$ peaks become progressively stronger and sharper, while the $S^{zz}$ peaks are smeared out and weakened. This trend is quantified by the extracted peak heights in Fig.~\ref{Fig3}(c). These results support the leading strong-coupling interpretation (Eq.~\eqref{eq:strongcoupling}) that the small-$\phi$ behavior is dominated by the DM-induced enhancement of the in-plane ordering.

We then examine whether the displacement field has more subtle effects beyond simply selecting the spin ordering plane and vector chirality. Motivated by the proximity of the CSL phase, we also measure the scalar chirality correlation function $\langle \chi_i\chi_j\rangle$ in the AFM phase at $U/t=12$ for representative $\phi$, as shown in Fig.~\ref{Fig3}(d). Rather unexpectedly, while the scalar chirality correlation decays rapidly at $\phi=0$, it develops a pronounced long-range behavior at finite displacement field $\phi\neq0$ (see the SM~\cite{supp} for additional evidence on $L_y=6$ cylinders). 
This observed long-range scalar chirality can originate either semiclassically from weak magnetic non-coplanarity, or purely quantum mechanically within a coplanar $120^\circ$-AFM state. 
From a symmetry point of view, such a scalar chirality would further break the composite symmetry of global spin-$z$ rotation and time-reversal, $\tilde{\mathcal T} = e^{-i\pi S^z_\text{tot}} \mathcal T$, preserved by the purely $xy$-plane $120^\circ$-AFM, rendering the finite-$\phi$ state a symmetry-distinct chiral AFM phase. 

\textit{Continuous CSL-AFM phase transition.}---
The persistence of scalar chirality across CSL and chiral AFM phases at finite $\phi$ has important implications for understanding the possible nature of the transition between them. 
We have examined the CSL-AFM transition at $\phi=0.005\pi$ in the SM~\cite{supp} and find that the transition appears to be continuous within our numerical resolution. 
Interestingly, a nearly continuous CSL-AFM transition was also suggested by a previous DMRG study at $\phi=0$~\cite{szaszChiralSpinLiquid2020}, despite the theoretical difficulty of realizing such a direct continuous transition when scalar chirality vanishes in the AFM phase.
Our finite-$\phi$ results substantially alter this scenario, as scalar chirality persists across the transition, naturally allowing several theoretically viable routes for continuity.
A particularly compelling scenario is a continuous spinon condensation transition within the chiral sector, which simultaneously develops magnetic order and destroys topological order while preserving chiral order. 
Such a transition is analogous to the previously studied $\nu=1/2$ bosonic Laughlin to superfluid transition~\cite{barkeshliContinuousTransitionFractional2014,songDeconfinedCriticalitiesDualities2023,songPhaseTransitionsOut2024a,wangEmergentQED$_3$Bosonic2026}, suggesting a possible description in terms of a $N_f=2$ QED$_3$ Chern-Simons theory.
Another alternative continuous scenario is a conventional magnetic transition, which would instead imply that the chiral AFM retains the topological order of the CSL and naturally belong to the 3D-XY universality class. 
This motivates future efforts to distinguish between these scenarios, potentially establishing twisted TMD homobilayers as a promising platform for studying fractionalized quantum criticality.

\begin{figure}[t]
    \centering
    \includegraphics[width=\linewidth]{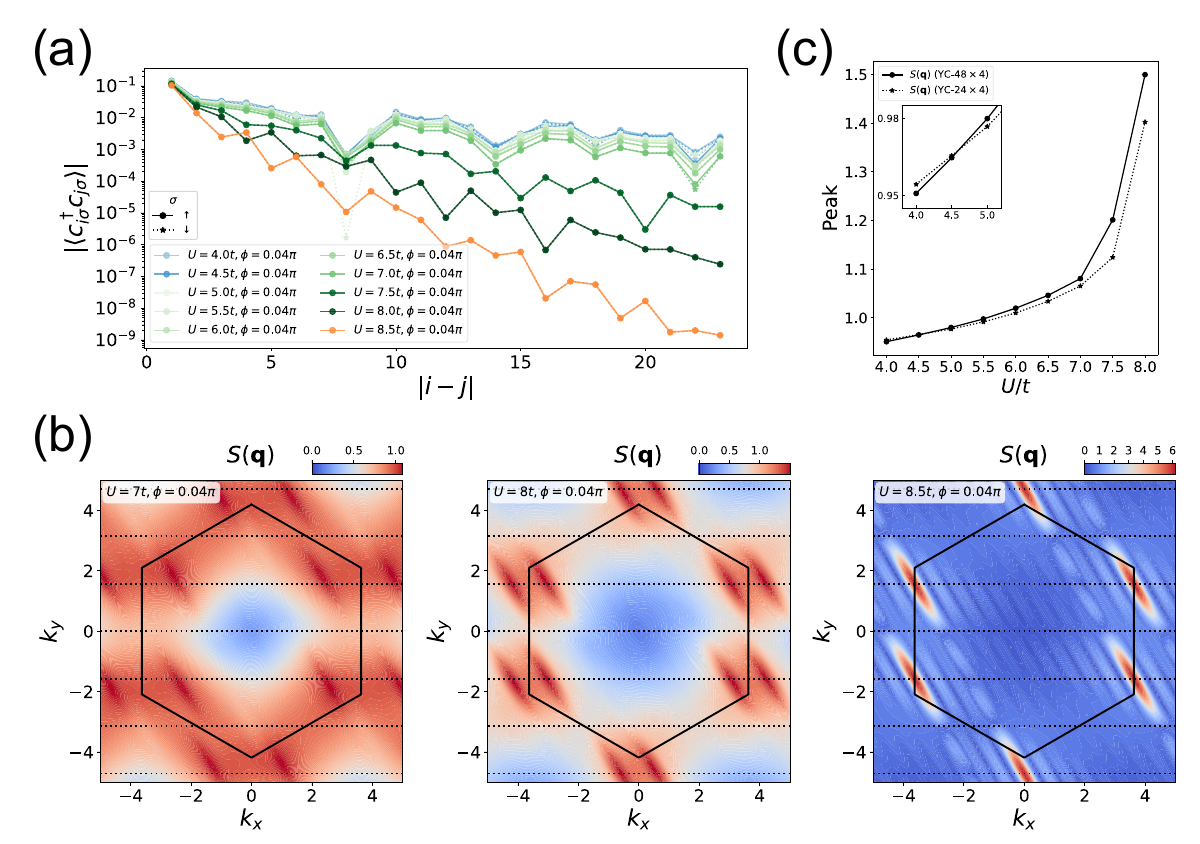}
    \caption{Metal-insulator transition and IC-SDW phases at finite displacement field $\phi=0.04\pi$ for different interaction strengths $U/t$ on $L_y=4$ cylinders. (a) Single-particle Green's function $\langle c^\dagger_{i\sigma} c_{j\sigma} \rangle$. (b) Spin structure factor $S(\mathbf{q})$ at representative values of $U/t=7$, $8$, and $8.5$. (c) Peak values of $S(\bm q)$ as a function of $U/t$ for two different system sizes.} 
    \label{Fig4}
\end{figure}

\textit{Incommensurate spin-density wave.}---
Having established the nature of large $U/t$ magnetic order, we now examine the metal-insulator transition at intermediate $U/t$ and finite $\phi$, where we find that IC-SDW phases can be stabilized. The single-particle Green's function $\langle c^\dagger_{i\sigma} c_{j\sigma}\rangle$ for various $U/t$ at fixed $\phi=0.04\pi$ is shown in Fig.~\ref{Fig4}(a). For $U/t \leq 7$, the Green's function decays slowly with distance and remains smoothly connected to the weak coupling regime, supporting a metallic ground state. Upon increasing $U/t \geq 7.5$, the Green's function undergoes a qualitative change to rapid exponential decay, signaling a metal-insulator transition. Fig.~\ref{Fig4}(b) compares the spin structure factors $S(\mathbf{q}) = \frac{1}{2} [S^{+-}(\mathbf{q})+ S^{-+}(\mathbf{q})] + S^{zz}(\mathbf{q})$ at representative interaction strengths $U/t=7$, $8$, and $8.5$. At $U/t\geq 8.5$, $S(\mathbf{q})$ shows a sharp single peak at the allowed momentum closest to $K$, consistent with the strong-coupling AFM phase identified previously. In contrast, for $U/t \leq 8$, the dominant peak of $S(\mathbf{q})$ splits into two peaks displaced from this momentum, characteristic of the emergence of IC-SDW order. Taken together, these results provide clear evidence on $L_y=4$ cylinders for successive IC-SDW metal, IC-SDW insulator, and strong-coupling AFM phases upon increasing $U/t$ at $\phi=0.04\pi$.

The formation of IC-SDW order can be naturally understood from a weak-coupling perspective. In Fig.~\ref{Fig1}(c), the spin-up and spin-down Fermi surfaces at finite $\phi$ can be approximately connected by nesting wave vectors close to $K$, which enhance the SDW susceptibility near the corresponding incommensurate momenta. However, since the nesting is non-perfect at generic $\phi$, IC-SDW order requires a finite interaction strength to develop from the metallic state, and the system becomes insulating only when the order is strong enough to fully gap the Fermi surface. We estimate the critical interaction for the metal to IC-SDW metal transition at $\phi=0.04\pi$ to be $U/t \sim 4.5$, based on the system size dependence of the spin structure factor peak, as shown in Fig.~\ref{Fig4}(c). 
At larger $U/t$, the stability of the IC-SDW order depends on detailed energetic competition.
Our results on $L_y=6$ cylinders~\cite{supp} instead suggest that the strong-coupling commensurate AFM phase becomes more favorable and may preempt the IC-SDW insulator phase. Thus, the IC-SDW insulator phase on the $L_y=4$ cylinder should be better viewed as a close competitor to the strong-coupling AFM phase, and further considering a small next-nearest-neighbor hopping could easily tilt the balance between them~\cite{kieseTMDsPlatformSpin2022}.

\textit{Discussion.}---
Our DMRG results on the minimal moiré Hubbard model have broad implications for moiré TMD systems. 
First, our results suggest that scalar chirality need not be confined to the CSL regime, but appears to be more robust and persists into the strong-coupling AFM phase at finite displacement field. 
The scalar chirality can be inferred from the composite time-reversal symmetry $\tilde {\mathcal T}$ breaking, making the magneto-optical Kerr effect~\cite{kerrXLIIIRotationPlane1877,argyresTheoryFaradayKerr1955,qiuSurfaceMagnetoopticKerr2000,xiaHighResolutionPolar2006,kapitulnikPolarKerrEffect2009} and thermal Hall transport~\cite{katsuraTheoryThermalHall2010,onoseObservationMagnonHall2010} natural experimental probes.
Displacement field evolution of these responses could reveal the onset of scalar chirality across the transition from conventional $120^\circ$-AFM to chiral AFM, and thereby pin down the critical field of the transition beyond our current numerical resolution. Temperature dependence can also be informative, since the scalar chirality order is expected to melt upon increasing temperature.

Second, although superconductivity (SC) is not directly resolved in our simulations, our results capture essential correlated physics and provide a useful context for discussing SC observed in twisted bilayer WSe$_2$ near~\cite{guoSuperconductivity50degTwisted2025,xiaBandwidthtunedMottTransition2026} or even directly at~\cite{xiaSuperconductivityTwistedBilayer2025,xiaBandwidthtunedMottTransition2026} half-filling proximate to the correlated insulating regime.
From a weak-coupling perspective, asymptotically exact renormalization group analysis of the moiré Hubbard model suggests a leading $d+id$ SC instability in the metal phase~\cite{wuPairDensityWaveChiralSuperconductivity2023,raghuSuperconductivityRepulsiveHubbard2010}, which may persist toward stronger coupling through enhanced quantum fluctuations~\cite{zhuSuperconductivityTwistedTransition2025,christosApproximateSymmetriesInsulators2025a,xieSuperconductivityTwistedWSe2025,guerciTopologicalSuperconductivityRepulsive2024a,tuoTheoryTopologicalSuperconductivity2025,qinTopologicalChiralSuperconductivity2025,fischerTheoryIntervalleycoherentAFM2025,jinGossamerSuperconductivityMoire2026}. 
Another interesting regime is the IC-SDW metal phase, where residual low-energy carriers may still support SC instability coexisting with magnetic order, as already observed in 5$^\circ$ twisted WSe$_2$~\cite{guoSuperconductivity50degTwisted2025}.
We further remark that, while previous DMRG studies have reported evidence of SC away from half-filling in moiré Hubbard and related models~\cite{wietekTunableStripeOrder2022,chenSingletTripletPair2023,biborskiSignaturesSuperconductingPairing2025}, directly resolving SC at half-filling with unbiased numerics remains an important open question.

Third, a recent intriguing experiment on twisted bilayer WSe$_2$~\cite{newexp} has revealed a rich internal structure within the Mott insulating regime, with an extended non-magnetic insulator emerging between SC and AFM insulator as the displacement field is increased.
Remarkably, this experimental phase sequence closely mirrors our results at $U/t\sim 8$--$9$, where the system evolves from a metal with SC instability to a CSL and then into a chiral AFM, supporting that the intermediate non-magnetic insulator may realize a CSL. 
It would therefore be particularly important for experiments to determine whether scalar chirality is present in both the candidate CSL and AFM regimes and whether the transition between them can be continuous.
We believe our work provides a key step toward understanding correlated phases in moiré TMDs and will stimulate further theoretical and experimental explorations.

\textit{Acknowledgements.}---
We would like to thank Steve Kivelson, Ming-Rui Li, Kin Fai Mak, and Jie Shan for helpful discussions. This work is supported in part by the NSFC under Grant Nos. 12347107 (C.T. and H.Y.) and 12334003 (H.Y.), and by the New Cornerstone Science Foundation through the Xplorer Prize (H.Y.).

\bibliography{ref.bib}

\end{document}


\title{Supplemental material for ``Chiral spin liquid and chiral antiferromagnetism in half-filled moiré Hubbard model: possible applications to twisted bilayer TMDs" }

\author{Chuyi Tuo}
\affiliation{Institute for Advanced Study, Tsinghua University, Beijing 100084, China}

\author{Hong Yao}

\affiliation{Institute for Advanced Study, Tsinghua University, Beijing 100084, China}

\date{\today}

\maketitle

\section{Chiral spin liquid to chiral antiferromagnet phase transition}

\begin{figure}[H]
    \centering
    \includegraphics[width=0.75\linewidth]{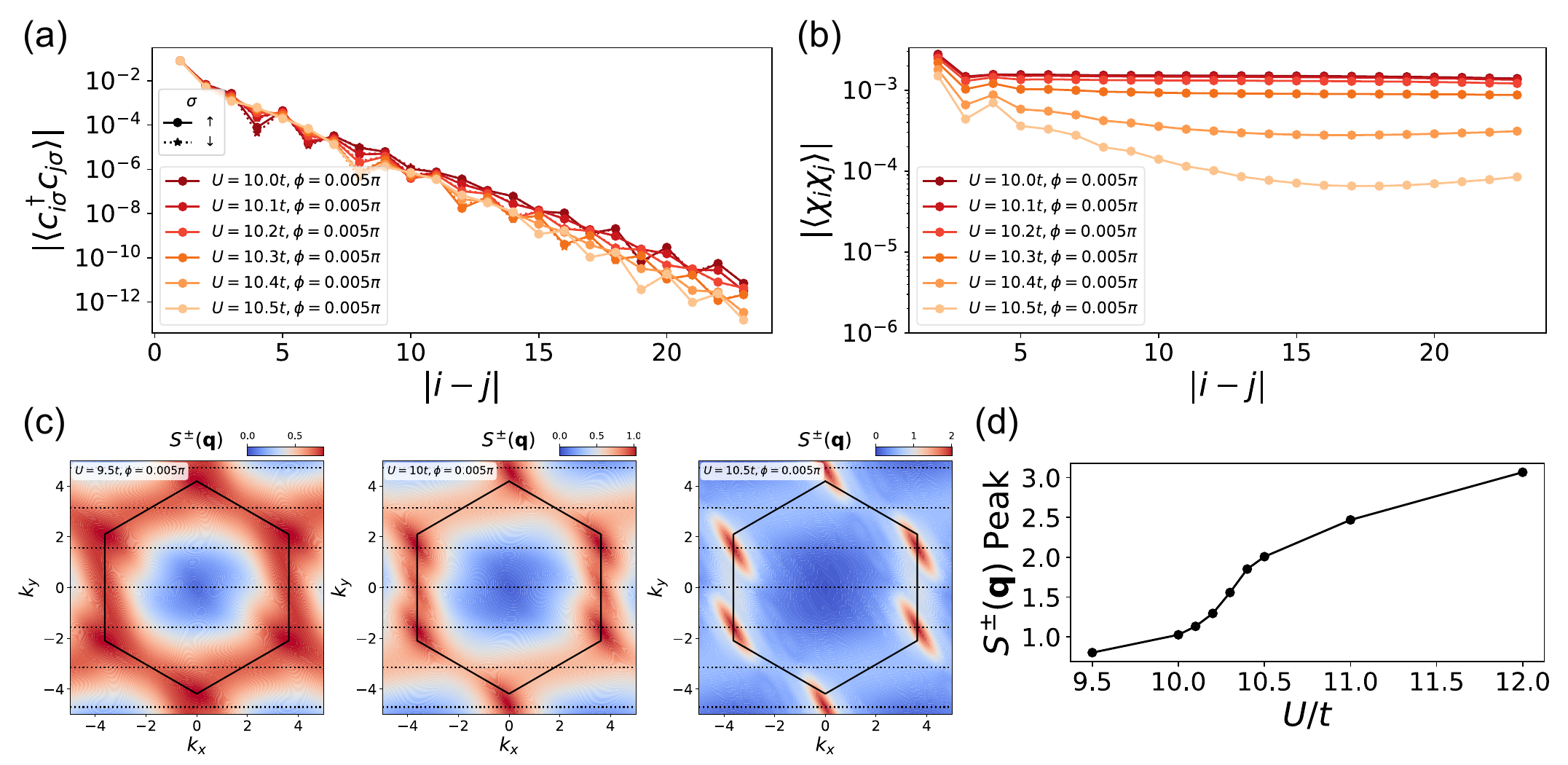}
    \caption{Characterization of the CSL to chiral AFM phase transition at finite displacement field $\phi=0.005\pi$ on $L_y=4$ cylinders. (a) Single-particle Green's function $\langle c^\dagger_{i\sigma} c_{j\sigma} \rangle$. (b) Scalar chirality correlation $\langle \chi_i \chi_j \rangle$. (c) Spin structure factor $S^{+-}(\mathbf{q})$ at representative interaction strengths $U/t=9.5$, $10$ and $10.5$. (d) Peak value of $S^{+-}(\mathbf{q})$ as a function of $U/t$. }
    \label{FigSupp_transition}
\end{figure}

In Fig.~\ref{FigSupp_transition}, we provide additional numerical results for the CSL to chiral AFM phase transition at a representative displacement field $\phi=0.005\pi$ on $L_y=4$ cylinders, and find that the transition appears to be continuous within our numerical resolution. 
As shown in Fig.~\ref{FigSupp_transition}(a), the single-particle Green's function $\langle c^\dagger_{i\sigma} c_{j\sigma} \rangle$ remains exponentially decay and almost unchanged across the transition, indicating a gapped single-particle sector that remains largely inert.
Turning to the chiral sector, Fig.~\ref{FigSupp_transition}(b) shows that the long-range scalar chirality correlation $\langle \chi_i \chi_j \rangle$ persists throughout the transition. Although the strength of the chiral order slightly decreases upon entering the chiral AFM phase at $U/t\gtrsim 10.3$ (see Fig.~\ref{FigSupp_transition}(d) for the estimation of phase boundary), the evolution remains smooth without any apparent discontinuity. 
The spin structure factors $S^{+-}(\mathbf q)$ in Fig.~\ref{FigSupp_transition}(c) provide further insight into how magnetic correlations evolve across the transition.
Specifically, $S^{+-}(\mathbf q)$ evolves smoothly from a broad continuum at $U/t=9.5$ deep in the CSL phase to an incipient $K$-point response at $U/t=10$ close to the CSL boundary, before the sharp $K$-point peaks dominate at $U/t=10.5$ in the chiral AFM phase. This precursor enhancement of $K$-point correlations within the CSL phase is suggestive of a continuous magnetic transition. 
A more quantitative view is provided by Fig.~\ref{FigSupp_transition}(d), where the peak of $S^{+-}(\mathbf q)$ increases smoothly as the system enters the chiral AFM phase, again showing no any visible discontinuity.
Although our numerical evidence consistently favors a continuous scenario, we emphasize that, within our current numerical resolution, we cannot exclude the possibility that the transition is weakly first order or involves a narrow intermediate phase.
Further numerical or experimental efforts to determine whether this transition is truly continuous at finite $\phi$ would be especially crucial for clarifying its nature and could open new avenues for exploring unconventional phases and phase transitions in moiré TMD systems.

\section{Additional numerical evidence on $L_y=6$ cylinders}
\subsection{DMRG setup on $L_y=6$ cylinders}
To further address finite circumference effects, we perform additional calculations on $L_y=6$ cylinders, which also have the advantage of directly accommodating the $K$ points associated with $120^\circ$-AFM order. Similar to the $L_y=4$ calculations in the main text, we adopt YC cylinder geometry, explicitly preserve $U(1)_\text{charge}\times U(1)_\text{spin}$ symmetry, and perform all simulations with bond dimension $D=10000$. Since the $L_y=6$ systems are substantially more demanding numerically for DMRG simulations, we restrict the cylinder length to $L_x=24$ to facilitate convergence.
The insulating phases generally show good convergence, with typical truncation errors of order $10^{-6}$. For the metallic phases, the stronger entanglement leads to larger truncation errors of order $10^{-4}$, but the calculations still allow us to identify the key qualitative features.

\subsection{Strong-coupling chiral antiferromagnetism}

\begin{figure}[H]
    \centering
    \includegraphics[width=0.75\linewidth]{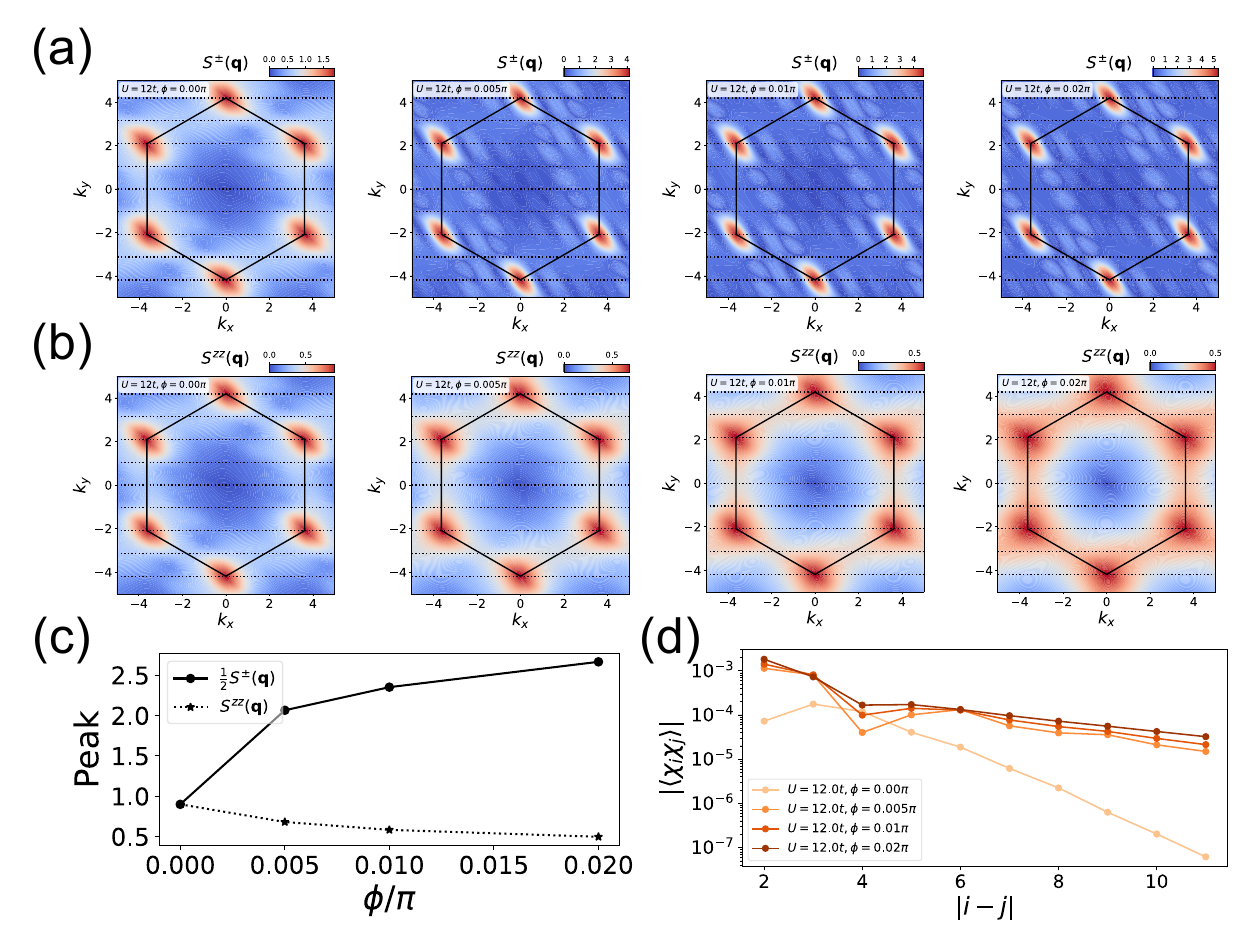}
    \caption{Characterization of the strong-coupling chiral AFM phase at $U/t=12$ for different displacement fields $\phi$ on $L_y=6$ cylinders. (a, b) Spin structure factors $S^{+-}(\mathbf{q})$ and $S^{zz}(\mathbf{q})$. (c) Peak values of $S^{+-}(\mathbf{q})$ and $S^{zz}(\mathbf{q})$ as a function of $\phi$. (d) Scalar chirality correlation $\langle \chi_i \chi_j \rangle$.}
    \label{FigSupp_AFM6}
\end{figure}

Fig.~\ref{FigSupp_AFM6} presents our DMRG characterization of the magnetic properties at $U/t=12$ on $L_y=6$ cylinders, closely reproducing the behavior observed on $L_y=4$ cylinders in Fig.~3 of the main text. Fig.~\ref{FigSupp_AFM6}(a) and (b) show the static spin structure factors $S^{+-}(\mathbf q)$ and $S^{zz}(\mathbf q)$, where the dominant magnetic peaks now occur precisely at the $K$ points. Fig.~\ref{FigSupp_AFM6}(c) further reveals a clear enhancement of the $S^{+-}(\mathbf q)$ peaks and suppression of the $S^{zz}(\mathbf q)$ peaks with increasing displacement field $\phi$. 
Moreover, as illustrated in Fig.~\ref{FigSupp_AFM6}(d), a finite displacement field $\phi$ also induces a pronounced enhancement of the scalar chirality correlation $\langle \chi_i \chi_j \rangle$ relative to the zero-field case $\phi=0$, consistent with the trend observed on $L_y=4$ cylinders. 
Although the accessible $L_y=6$ cylinder length is not yet sufficient to resolve a well-developed long-distance correlation plateau, owing to the relatively long chiral correlation length, the correlation magnitude over comparable distances remains similar to that found on $L_y=4$ cylinders.
Taken together, the close agreement between the $L_y=4$ and $L_y=6$ results indicates that the chiral behavior remains robust against increasing cylinder circumference and supports the possibility that the chiral order persists toward the two-dimensional limit.

\subsection{Incommensurate spin-density wave}

\begin{figure}[H]
    \centering
    \includegraphics[width=\linewidth]{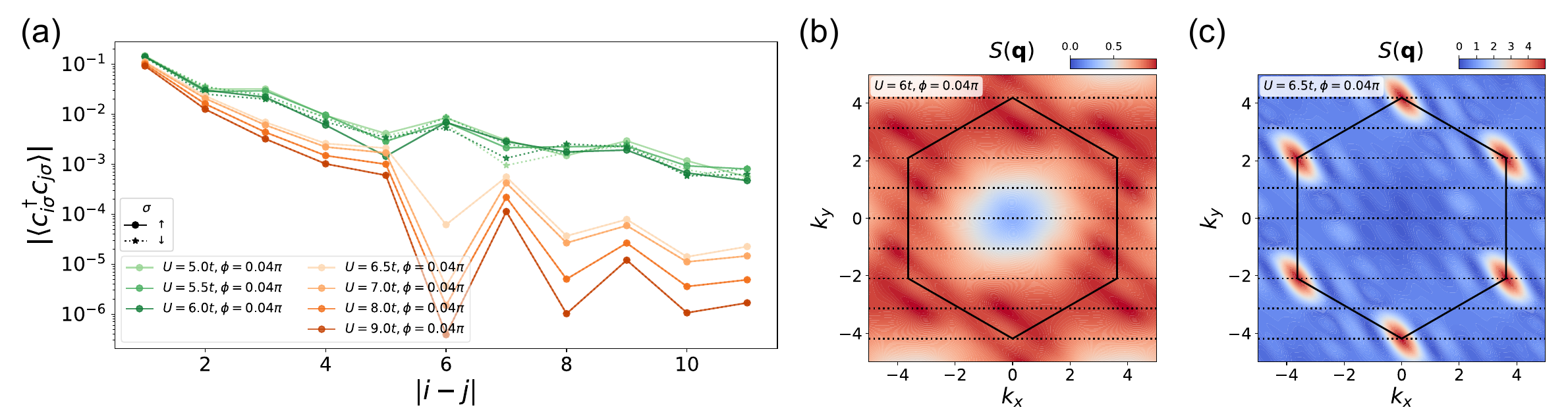}
    \caption{Metal-insulator transition and IC-SDW phase at finite displacement field $\phi=0.04\pi$ for different interaction strengths $U/t$ on $L_y=6$ cylinders. (a) Single-particle Green's function $\langle c^\dagger_{i\sigma} c_{j\sigma} \rangle$. (b, c) Spin structure factor $S(\mathbf{q})$ at $U/t=6$ and $6.5$.}
    \label{FigSupp_ICSDW6}
\end{figure}

Fig.~\ref{FigSupp_ICSDW6} examines the metal-insulator transition and IC-SDW behavior at $\phi=0.04\pi$ on $L_y=6$ cylinders, where we find that the detailed phase sequence is modified by the energetic competition with the strong-coupling commensurate AFM. As shown in Fig.~\ref{FigSupp_ICSDW6}(a), the single-particle Green's function $\langle c^\dagger_{i\sigma} c_{j\sigma} \rangle$ decays slowly for $U/t\leq6$ but exhibits a rapid exponential decay for $U/t\geq6.5$, indicating a metal-insulator transition between $U/t=6$ and $6.5$.
On the metallic side close to the transition, Fig.~\ref{FigSupp_ICSDW6}(b) shows the spin structure factor $S(\mathbf{q})$ at $U/t=6$, which remains broad but exhibits a clear incommensurate magnetic response characteristic of an IC-SDW metal phase. Despite the more challenging convergence of the $L_y=6$ metallic phase, the resulting IC-SDW response shows a clearly more two-dimensional character than that observed on $L_y=4$ cylinders.
On the insulating side close to the transition, Fig.~\ref{FigSupp_ICSDW6}(c) reveals that the spin structure factor $S(\mathbf{q})$ is already dominated by $K$-point peaks at $U/t=6.5$, suggesting a direct transition into the commensurate AFM on $L_y=6$, with no intervening IC-SDW insulator phase as found on $L_y=4$ cylinders.
This behavior can be naturally understood from the compatibility of the $L_y=6$ geometry with the $K$ points, which favors the strong-coupling commensurate AFM and allows it to prevail before the IC-SDW becomes strong enough to fully gap the Fermi surface.
Therefore, as noted in the main text, the IC-SDW insulator phase observed on $L_y=4$ cylinders is better viewed as a close competitor to the strong-coupling commensurate AFM, with their energetic balance readily shifted by experimentally relevant perturbations such as a small next-nearest-neighbor hopping.

